\documentclass[a4paper,fleqn,usenatbib]{mnras}

\usepackage{newtxtext,newtxmath}

\usepackage[T1]{fontenc}
\usepackage{ae,aecompl}
\usepackage{courier}

\usepackage{graphicx}	
\usepackage{amsmath}	
\usepackage{amssymb}	

\title[Environment and the AGN Fundamental Plane]{The role of environment in the observed Fundamental Plane of radio Active Galactic Nuclei}

\author[Stanislav S. Shabala]{
Stanislav S. Shabala$^{1}$\thanks{E-mail: stanislav.shabala@utas.edu.au}
\\
$^{1}$ School of Natural Sciences, Private Bag 37, University of Tasmania, Hobart, TAS 7001, Australia
}

\date{Accepted XXX. Received YYY; in original form ZZZ}

\pubyear{2017}

\begin{document}
\label{firstpage}
\pagerange{\pageref{firstpage}--\pageref{lastpage}}
\maketitle

\begin{abstract}

The optical Fundamental Plane of black hole activity relates radio continuum luminosity of Active Galactic Nuclei to [O\,{\textsc{iii}}] luminosity and black hole mass. We examine the environments of low redshift ($z<0.2$) radio-selected AGN, quantified through galaxy clustering, and find that halo mass provides similar mass scalings to black hole mass in the Fundamental Plane relations. AGN properties are strongly environment-dependent: massive haloes are more likely to host radiatively inefficient (low-excitation) radio AGN, as well as a higher fraction of radio luminous, extended sources. These AGN populations have different radio -- optical luminosity scaling relations, and the observed mass scalings in the parent AGN sample are built up by combining populations preferentially residing in different environments. Accounting for environment-driven selection effects, the optical Fundamental Plane of supermassive black holes is likely to be mass-independent, as predicted by models.

\end{abstract}


\section{Introduction}

The Fundamental Plane (FP) of black hole activity defines a low-redshift correlation between radio continuum luminosity, X-ray luminosity and black hole mass of the form $L_{\rm radio} \propto L_X^{a} M^{b}$. It holds over 10 orders of magnitude in radio luminosity, and connects Active Galactic Nuclei (AGN) hosted by supermassive black holes with (much more radio-quiet) X-ray black hole binaries (XRBs). Work by various authors \citep{MerloniEA03,KoerdingEA06,GultekinEA09,BonchiEA13,NisbetBest16} has confirmed the existence of the FP, with exponents in the range $a=0.4-0.8$, $b=0.6-1.0$, depending on the sample used.

\citet{HeinzSunyaev03} pointed out that the expected scaling between X-ray and radio luminosity depends on whether the X-ray emission is dominated by the hot disk corona, or the jet itself through optically thin synchrotron or synchrotron self-Compton emission. \citet{FalckeEA04} showed that, for jet-dominated X-ray emission, the mass dependence of the FP naturally comes about due to the changing break frequency between the optically thin and thick parts of the jet. Typical spectral indices\footnote{We use the convention $F_{\nu} \propto \nu^{\alpha}$ throughout this paper.} of $\alpha=+0.1$ and $-0.6$ for the optically thick and thin parts of the SED, respectively, yield $a=0.7$ and $b=0.6$, consistent with observed scalings in the low/hard state of black hole X-ray binaries \citep{KoerdingEA06,CorbelEA13}; similar scalings are obtained if the X-rays are produced by a radiatively inefficient accretion flow, rather than the jet \citep{MerloniEA03}. For a self-absorbed synchrotron jet with X-ray emission coming from the accretion disk, theory predicts a mass-{\it independent} scaling $L_{\rm radio} \propto L_X^{1.4}$  \citep{FalckeEA04}, consistent with observations of at least some radio-faint XRBs \citep[e.g. H1743-322; ][but see \citet{RushtonEA16}]{CoriatEA11} and Atoll-type neutron star X-ray binaries \citep{MigliariFender06,TetarenkoEA16}.

\citet{SaikiaEA15} recently reported the existence of the so-called optical FP, which uses the [O\,{\textsc{iii}}] narrow-line luminosity instead of X-ray luminosity. Line emission from the narrow-line region scales with the strength of the ionising radiation field, and has been shown to trace AGN bolometric luminosity \citep{HeckmanEA04}; hence it is a good proxy for the black hole accretion rate. By converting the measured [O\,{\textsc{iii}}] luminosities to hard (2--10 keV) X-ray luminosities, \citet{SaikiaEA15} compared their extrapolated FP for 39 AGN (12 Seyferts, 20 LINERs, 7 transition objects) to observations of XRBs, finding that the observed XRB population has higher radio luminosities than expected from the AGN-only FP. This result, which brings into question the FP unification of supermassive and stellar mass black holes, has since been confirmed using a sample of lower-luminosity AGN (LINERs) by \citet{NisbetBest16}.
 
In this paper, we suggest that the observed mass dependence in the optical FP relations for supermassive black holes is due to environmental effects. High and low-mass radio AGN preferentially inhabit different environments. As a consequence, these AGN are found in different accretion modes, and have different scalings between jet kinetic power and radio continuum luminosity.

\section{Sample}
\label{sec:sample}

The starting point for our sample of radio AGN comes from \citet{BestHeckman12}. These authors cross-matched the seventh data release of the SDSS spectroscopic sample \citep{AbazajianEA09} with the NVSS \citep{CondonEA98} and FIRST \citep{BeckerEA95} 1.4 GHz radio continuum surveys. \citet{BestHeckman12} used a combination of indicators to identify radio AGN, including the 4000\AA\, break, ratio of radio luminosity to stellar mass, and optical emission line diagnostics. The \citet{BestHeckman12} sample consists of 7302 radio AGN with redshifts $0.01 \leq z \leq 0.3$ and host galaxy {\it r}-band magnitude brighter than 17.77. These authors further separated radio AGN into High (HERG) and Low (LERG) Excitation Radio Galaxies, based on optical line ratios of six species, as well as the [O\,III] line equivalent width. Although only around one third of the objects in the \citet{BestHeckman12} sample had sufficiently strong emission lines to be classified spectroscopically, the majority of unclassified sources were found towards the back of the sample volume. In the analysis below, we restrict the AGN sample to $z < 0.2$. Early theoretical \citep{MerloniEA03} and observational \citep{KoerdingEA06} work on the Fundamental Plane explicitly excluded radiatively efficient AGN, however much subsequent work \citep[e.g. ][]{GultekinEA09,BonchiEA13,SaikiaEA15} has employed samples containing both high and low-excitation radio galaxies. As shown in Section~\ref{sec:results}, this introduces a systematic bias in the derived relations.

To study the correlation between radio and narrow line luminosities, we need to construct a complete subsample. NVSS is 99 percent complete at 3.4\,mJy \citep{CondonEA98}. Above a narrow-line luminosity integrated flux of $S_{\rm [O\,{\textsc{iii}}]} > 20 \times 10^{-17}$ ergs/s/cm$^2$, we found 95\% of the AGN have a signal-to-noise ratio of at least two. Applying these two cuts to the \citet{BestHeckman12} sample yields 1356 LERGs and 138 HERGs at $z<0.2$.

To quantify environments, we use the SDSS group catalogues of \citet{YangEA08}. These authors used an adaptive halo-based group finder to associate nearby galaxies with a common dark matter halo, both in rich and poor systems. Tests with mock galaxy catalogues suggest a typical accuracy for the halo mass estimates of 0.3 dex. Cross-matching with the \citet{YangEA08} catalogue yields 836 AGN with halo masses. Of these, 726 are classified by \citet{BestHeckman12} as LERGs, 68 as HERGs, and the remaining 42 objects do not have an optical classification; we do not consider objects without a spectroscopic classification in our analysis. We further restrict our sample to sources which we can reliably classify as compact or extended (Section~\ref{sec:OIII_radio_relation}). Our final sample consists of 714 LERGs and 67 HERGs with halo masses.

\section{Theoretical considerations}
\label{sec:theory}

\subsection{Extended radio sources}

Analytical models of powerful double (Fanaroff-Riley type II; FR-II \citep{FR74}) radio sources \citep[e.g.][]{Scheuer74,KA97,TS15} describe supersonic expansion of radio lobes inflated by backflow of shocked plasma from the termination point of an initially relativistic jet. Conservation of energy within the cocoon and the shocked gas shell, together with the ram pressure condition at the hotspot, are sufficient for describing cocoon dynamics. Similar models apply to lobed lower-power FR-I sources. In jetted FR-Is, the initially conical jets are not collimated sufficiently by the external medium before stalling \citep{Alexander06,KrauseEA12}; beyond the stalling point the jet suffers significant entrainment, yielding diffuse plumes of radio emission. The synchrotron emission is usually calculated under the assumption of a constant fraction of the total lobe energy being in the magnetic field \citep{KDA97}. It can be shown \citep[e.g.][see their Eqn 4]{SG13} that, in the absence of loss processes\footnote{Equation~\ref{eqn:Lradio_Qjet_scaling} is evaluated under the assumption of no synchrotron or Inverse Compton losses in the emitting population. In reality, these losses become progressively more important as the source ages, and manifest themselves via a systematic shift in the radio luminosity -- jet power relation with increasing source size (see Figure 2 of \citet{SG13}).}, the relationship between monochromatic lobe luminosity $L_{\nu}$ and jet kinetic power $Q_{\rm jet}$ is
\begin{equation} \label{eqn:Lradio_Qjet_scaling}
L_{\rm \nu} = A_1 Q_{\rm jet}^{\frac{5+p}{6}} \left( \frac{\nu}{\nu_0} \right)^{\frac{1-p}{2}}
\end{equation}
Here, the initial electron Lorentz factor distribution at the site of acceleration is given by $n(\gamma) \propto  \gamma^{-p}$, and $A_1$ is a constant. The optically thin synchrotron spectral index is related to $p$ via $\alpha=(1-p)/2$.

Let jet power generation efficiency be $\eta$, i.e. $Q_{\rm jet}=\eta \dot{M}_{\rm BH} c^2$; and bolometric luminosity be $L_{\rm bol}=\epsilon_{\rm rad} \dot{M}_{\rm BH} c^2$. Then, $Q_{\rm jet} = \frac{\eta}{\epsilon_{\rm rad}} L_{\rm bol}$ and Equation~\ref{eqn:Lradio_Qjet_scaling} gives the relationship between radio and bolometric luminosities,
\begin{equation}\label{eqn:extended_radioBol_scaling}
\log ( L_{\rm \nu} ) = \left[  \log A_2 +   \left( {\frac{5+p}{6}} \right) \log \left( \frac{\eta}{\epsilon_{\rm rad}} \right)   \right]  +  \left( {\frac{5+p}{6}} \right) \log L_{\rm bol}
\end{equation}
where the frequency dependence has been absorbed into the constant $A_2$. For $p=2.6$ (corresponding to spectral index $\alpha=-0.8$) the slope of the $\log L_{\rm \nu}$ -- $ \log L_{\rm bol}$ relation is 1.27.

\subsection{Compact sources}
\label{sec:compact_theory}

\citet{FalckeBiermann95} and \citet{HeinzSunyaev03} derived a general expression for the integrated flux density of a self-similar jet. The scaling between jet kinetic power and flux density (or radio luminosity) depends on the details of the jet model. For jet models in which the energy density in the magnetic field and relativistic particles both scale with total pressure, $L_{\nu} \propto Q_{\rm jet}^{17/12 + \alpha/3}$. For steep-spectrum jets ($\alpha=-0.8$) this gives an exponent of 1.15, or 1.42 for flat-spectrum jets with $\alpha \sim 0$. Synchrotron self-absorption, which naturally produces flat spectra \citep{BlandfordKonigl79}, takes place on parsec scales, and for typical jet powers the synchrotron emission should be optically thin on kiloparsec scales which are of interest in this work. On scales comparable to galactic disk thickness, the observed spectra of (intrinsically) optically thin jets can still show relatively flat or even peaked spectra due to free-free absorption by the interstellar medium \citep{BicknellEA97}. The assumption of proportionality between magnetic, particle and total pressure is also made in standard radio lobe models, and hence it is not surprising that for both extended lobes and compact jets we expect similar scalings between jet power (or a proxy for this quantity, such as bolometric luminosity) and synchrotron radio luminosity. For gas-dominated disks the scaling is slightly different \citep{HeinzSunyaev03}, $L_{\nu} \propto Q_{\rm jet}^{1.65 + 0.45 \alpha}$; for $\alpha=-0.8$ this yields a slope of $1.29$, similar to the above values for both compact and extended sources.

\subsection{Comparing compact and extended emission}
\label{sec:theory_summary}

As shown above, both compact and extended steep-spectrum sources are expected to follow $L_{\rm radio} \propto L_{\rm bol}^x$ for $x=1.15-1.42$. The main difference is in the normalization: in a large, low-redshift ($z<0.3$) sample of powerful radio galaxies, \citet{HardcastleEA98} find the ratio of core to total luminosity at 178 MHz to be between $0.001$ and $0.4$, with the median value of $0.02$. We therefore expect compact sources to have substantially lower radio luminosities at a given bolometric luminosity.

There are two further noteworthy points. First, the above discussion implicitly assumes that core and extended radio flux relate to the same jet kinetic power. However, these two measurements probe different timescales: extended radio luminosity is a proxy for time-averaged jet power on timescales of tens to hundreds of Myrs, while core luminosity traces the (quasi-)instantaneous jet power. For variable accretion rate these two measurements will in general be different, although they should converge to similar values for a continuously active jet, i.e. the time-averaged jet power should be similar the average instantaneous power. In the case of intermittent jet activity, the jet power inferred from extended radio luminosity will be lower by a factor corresponding to the jet duty cycle. The majority of radio sources in the local Universe are hosted by massive galaxies with high duty cycles \citep{BestEA05}, hence we do not expect this effect to be important to the present work.

Second, recognising the need to probe nuclear emission, a number of authors \citep[e.g. ][]{MerloniEA03,KoerdingEA06,GultekinEA09} have used compact (arcsec-scale) 5 GHz or even 15 GHz \citep{SaikiaEA15} radio flux densities. However, for all but the nearest AGN these observations still probe scales of several kpc; the problem is exacerbated at higher redshifts \citep[e.g. ][]{BonchiEA13}. Below, we show that the contribution of extended flux is environment-dependent.


\begin{figure}
	\includegraphics[width=1.0\columnwidth]{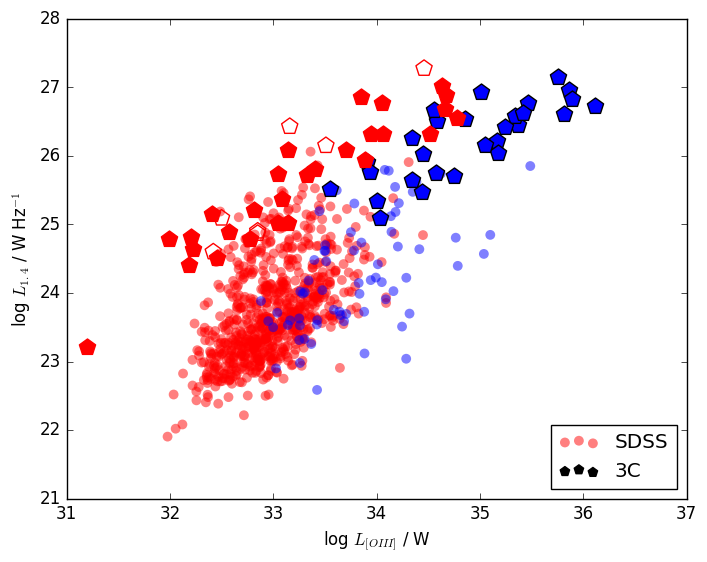}
	\caption{1.4 GHz radio luminosity - [O\,{\textsc{iii}}] line luminosity correlation. Circles are the complete low-luminosity \citet{BestHeckman12} sample with halo masses. Pentagons are powerful 3C radio sources from \citep{ButtiglioneEA10}; filled symbols denote an [O\,{\textsc{iii}}] detection, open symbols are upper limits on [O\,{\textsc{iii}}] luminosity. High-excitation sources (blue) are offset from the low-excitation (red) population. Powerful extended radio galaxies clearly trace out the upper (high radio luminosity) envelope of the underlying NVSS population.}
	\label{fig:Lnvss_vs_Loiii_complete}
\end{figure}

\begin{figure*}
	\includegraphics[width=1.0\columnwidth]{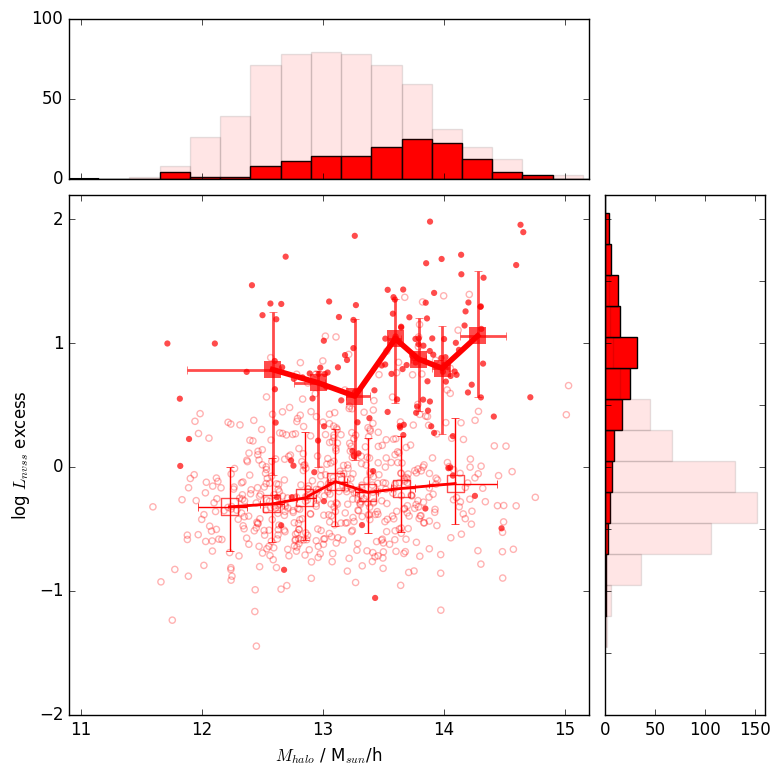}
	\includegraphics[width=1.0\columnwidth]{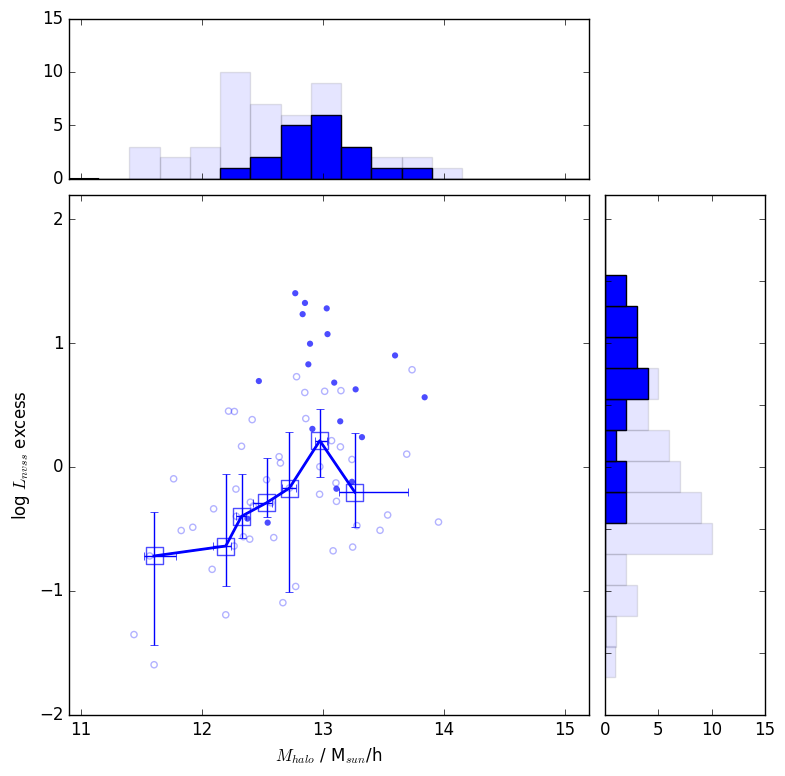}
	\caption{Residuals to the radio - narrow line luminosity correlation, as a function of halo mass, plotted separately for LERGs ({\it left}) and HERGs ({\it right} panel). Open circles correspond to compact sources, closed circles are extended objects. Large symbols show medians of the binned populations, with each bin chosen to have the same number of objects; error bars contain 68 percent of the data in each halo mass bin. There are only 19 extended HERGs, and hence no binning is shown for these objects. In both HERGs and LERGs, extended sources clearly have higher radio luminosities, and preferentially reside in denser environments.}
	\label{fig:Lnvss_vs_Loiii_resids}
\end{figure*}

\section{The optical Fundamental Plane}
\label{sec:results}

\subsection{Optical -- radio luminosity relation}
\label{sec:OIII_radio_relation}

The left panel of Figure~\ref{fig:Lnvss_vs_Loiii_complete} shows the relationship between [O\,{\textsc{iii}}]\footnote{Our [O\,{\textsc{iii}}] luminosities are not extinction-corrected, consistent with the luminosity-independent bolometric correction reported by \citet{HeckmanEA04}. \citet{SaikiaEA15} showed that using dust-corrected [O\,{\textsc{iii}}] luminosity results in lower $a_{\rm OIII}$ and higher $a_{\rm mass}$ values (see Table~\ref{tab:fit_params}) by 0.3-0.4 dex, due to the luminosity dependence of the bolometric correction \citep{LamastraEA09}.} and radio continuum luminosity in our sample. High and low-excitation radio AGN both show a correlation, as previously reported by other authors \citep{WillottEA99,ButtiglioneEA10}. The different normalizations for HERG and LERG populations are expected due to their different jet production efficiencies $\eta / \epsilon_{\rm rad}$ (Equation~\ref{eqn:extended_radioBol_scaling}). The low radio luminosity end is dominated by LERGs, while number counts of LERGs and HERGs become comparable at the highest ($L_{\rm 1.4}>10^{25}$~W/Hz) luminosities (Figure~\ref{fig:Lnvss_vs_Loiii_complete}; see also Figure 4 of \citet{BestHeckman12}). For comparison, we overplot the sample of extended 3C radio galaxies of \citet{ButtiglioneEA10}, who also used spectroscopic diagnostics to classify AGN as High or Low-Excitation. Extended radio galaxies follow similar trends to the lower-luminosity sample, but clearly occupy the upper envelope of the distribution; in other words, they have high radio luminosities relative to their [O\,{\textsc{iii}}] luminosity.

It is instructive to consider the residuals to the best-fit relation between these variables. Figure~\ref{fig:Lnvss_vs_Loiii_resids} shows the excess in radio luminosity (traced by NVSS) as a function of halo mass, plotted separately for LERGs and HERGs. Here, the radio excess has been calculated by subtracting from each source the best-fit to the appropriate $L_{\rm 1.4} - L_{\rm [O\,{\textsc{iii}}]}$ relation. Each population has further been split into compact and extended sources, using the classifications of \citet{BestHeckman12}. These authors classified sources into four categories: (1) those sources which consist of single components in both FIRST and NVSS; (2) sources with a single NVSS component but multiple components in FIRST; (3) NVSS sources without a FIRST counterpart; and (4) NVSS sources with multiple components. We classify sources in categories (2) and (4) as extended. We classify category (1) sources as extended if their NVSS to FIRST flux density ratio exceeds 5, i.e. they exhibit significant low surface brightness emission which is resolved out by FIRST. Category (1) sources in which this flux density ratio is below the threshold are classified as compact. This approach follows \citet{KimballIvezic08,SinghEA15}, albeit with a higher value of the flux density ratio (those authors adopt a threshold value of 1.4); our results are qualitatively similar for lower values of the flux density ratio threshold. Finally, we exclude category (3) objects from our analysis, noting that only 1 HERG and 12 LERGs fall into this category. There are 138 extended and 576 compact LERGs; and 19 extended, 48 compact HERGs. There is no statistically significant (as given by a Kolmogorov-Smirnov test at the 10 percent level) difference in the redshift distributions of compact and extended sources. As a caveat, we note that compact sources defined using 1.4 GHz flux densities may encompass more non-jetted emission than higher frequency samples \citep[e.g. ][]{KoerdingEA06,GultekinEA09,SaikiaEA15}.


Three features are immediately apparent in Figure~\ref{fig:Lnvss_vs_Loiii_resids}. First, High-Excitation Radio Galaxies preferentially reside in poorer environments, with a median halo mass of $6 \times 10^{12} M_{\odot}/h$ compared to $1.6 \times 10^{13} M_{\odot}/h$ for LERGs; this result echoes the findings of \citet{SabaterEA13}. Second, both HERGs and LERGs show a clear dichotomy between compact and extended sources, with extended sources consistently exhibiting radio emission in excess of the median value for the combined (compact plus extended) population. Third, compact and extended source counts depend on environment: despite compact sources greatly outnumbering extended sources in both HERGs and LERGs, the numbers are comparable at the highest masses (above $10^{13} M_{\odot}/h$ for HERGs, and $10^{14} M_{\odot}/h$ for LERGs). Hence, at a fixed [O\,{\textsc{iii}}] luminosity, massive haloes are more likely to host extended, high-luminosity radio sources. The most radio bright sources will, therefore, be preferentially found in rich environments. Because halo mass correlates with AGN host galaxy mass and therefore black hole mass, any mass-dependent scaling will push the extended, high-luminosity sources to the right in $L_{1.4} - L_{[O\,{\textsc{iii}}]}$ plane, thereby strengthening any existing correlation. The separation between HERGs and LERGs (Figure~\ref{fig:Lnvss_vs_Loiii_complete}) will also be reduced with a mass scaling, since HERGs preferentially inhabit poor environments; however a separate relation, corresponding to a different $\eta / \epsilon_{\rm rad}$ value from the LERG population, would be a more physically-motivated alternative.

\subsection{Thickness of the Fundamental Plane}
\label{sec:scatter}

A number of authors \citep[e.g.][]{MerloniEA03,Daly16} have suggested that parameters such as black hole spin add scatter to the Fundamental Plane. The above analysis suggests that environmental effects are also likely to contribute. Table~\ref{tab:fit_params} shows the results of fitting a plane in radio luminosity, [O\,III] luminosity and mass (either halo or black hole, estimated from velocity dispersion using the relation of \citet{TremaineEA02}\footnote{As shown by \citet{NisbetBest16}, using the steeper relation of \citet{McConnellMa13} will change $a_{\rm mass}$ by 0.2-0.3 dex.}) to our data, using the {\texttt{Hyperfit}}\footnote{hyperfit.icrar.org} package \citep{RobothamObreschkow15}. 
\texttt{Hyperfit} uses traditional likelihood methods to estimate a best-fitting model to multi-dimensional data, in the presence of parameter covariances, intrinsic scatter and heteroscedastic errors on individual data points. It assumes that both the intrinsic scatter and uncertainties on individual measurements are Gaussian, and allows for error covariance between orthogonal directions. Parameter estimates in Table~\ref{tab:fit_params} have been obtained using the Conjugate Gradients method; similar results are found using a quasi-Newtonian method.
 The scatter about the best-fit relation of $\sim 0.5$~dex is comparable with results reported by other authors \citep{MerloniEA03,BonchiEA13,SaikiaEA15,NisbetBest16}.

Compact sources show a marginally steeper dependence of radio luminosity on [O\,III] luminosity than extended sources, and substantially less scatter about this relation. This is expected from synchrotron ageing and jet-environment interaction in extended sources, discussed in Section~\ref{sec:theory}, and has also been reported by \citet{MiraghaeiBest17}. Compact and extended sources have different normalisations of the radio -- optical luminosity relation, and preferentially reside in different environments (Figure~\ref{fig:Lnvss_vs_Loiii_complete}). This likely accounts for the reduced scatter when mass is introduced as an additional variable: the mass scalings (either halo or black hole) of the combined population are steeper than for compact and extended sources individually; a similar point applies to combining the HERG and LERG samples. Radio - optical luminosity correlations depend only weakly on mass in individual subsamples, with pronounced mass scalings built up by combining different populations. A key result of this work is that the scatter in the Fundamental Plane which uses halo mass is the same as that using black hole mass. This is consistent with our interpretation that black hole mass is a proxy for environment.

Our radio -- [O\,{\textsc{iii}}] luminosity slope is consistent with the results of \citet{SaikiaEA15}. Both our relation and that of \citet{SaikiaEA15} are shallower than the slope predicted by Equation~\ref{eqn:extended_radioBol_scaling}, and also the observed scaling relations in powerful radio sources \citep{WillottEA99}. Different jet production efficiencies $\eta / \epsilon_{\rm rad}$ across the AGN populations will contribute to flattening this slope. Recently, \citet{SbarratoEA14} have argued that at the lowest accretion rates the ionising luminosity (traced by [O\,{\textsc{iii}}]) can drop significantly below a linear scaling. Such a decrease would lead to higher $L_{1.4} / L_{\rm [O\,{\textsc{iii}}]}$ ratios at the lowest [O\,{\textsc{iii}}] luminosities, and again flatten the observed relation.

\begin{table}
\centering
\tiny
\tabcolsep=0.08cm
\begin{tabular}{rccccccccc}
\hline

		&		&	No mass	&		&	$M_{\rm halo}$	&		&		&	$M_{\rm bh}$	&		&		\\
		&	$N_{\rm AGN}$	&	$a_{\rm OIII}$	&	$\sigma$	&	$a_{\rm OIII}$	&	$a_{\rm mass}$	&	$\sigma$	&	$a_{\rm OIII}$	&	$a_{\rm mass}$	&	$\sigma$	\\
\hline\\	
	{\bf All sources}	&	781	&$	0.85 \pm 0.05	$ & $	0.45	$ & $	0.87 \pm 0.05	$ & $	0.28 \pm 0.04	$ & $	0.42	$ & $	0.91 \pm 0.05	$ & $	0.25 \pm 0.05	$ & $	0.42	$ \\
	Compact	&	576	&$	0.90 \pm 0.04	$ & $	0.33	$ & $	0.91 \pm 0.04	$ & $	0.11 \pm 0.03	$ & $	0.32	$ & $	0.93 \pm 0.04	$ & $	0.18 \pm 0.03	$ & $	0.31	$ \\
	Extended	&	138	&$	0.67 \pm 0.10	$ & $	0.47	$ & $	0.73 \pm 0.10	$ & $	0.18 \pm 0.10	$ & $	0.45	$ & $	0.74 \pm 0.10	$ & $	0.20 \pm 0.11	$ & $	0.45	$ \\
	{\bf LERGs}: All	&	714	&$	0.94 \pm 0.06	$ & $	0.43	$ & $	0.93 \pm 0.06	$ & $	0.25 \pm 0.04	$ & $	0.40	$ & $	0.97 \pm 0.06	$ & $	0.16 \pm 0.06	$ & $	0.41	$ \\
	Compact	&	576	&$	1.02 \pm 0.05	$ & $	0.30	$ & $	1.03 \pm 0.05	$ & $	0.05 \pm 0.03	$ & $	0.29	$ & $	1.03 \pm 0.05	$ & $	0.12 \pm 0.04	$ & $	0.29	$ \\
	Extended	&	138	&$	0.88 \pm 0.15	$ & $	0.43	$ & $	0.93 \pm 0.14	$ & $	0.17 \pm 0.10	$ & $	0.40	$ & $	0.92 \pm 0.14	$ & $	0.02 \pm 0.13	$ & $	0.42	$ \\
	{\bf HERGs}: All	&	67	&$	0.80 \pm 0.16	$ & $	0.52	$ & $	0.60 \pm 0.97	$ & $	0.62 \pm 0.95	$ & $	0.45	$ & $	0.76 \pm 0.16	$ & $	0.55 \pm 0.14	$ & $	0.43	$ \\
	Compact	&	48	&$	0.62 \pm 0.18	$ & $	0.45	$ & $	0.34 \pm 0.27	$ & $	0.73 \pm 0.22	$ & $	0.37	$ & $	0.61 \pm 0.16	$ & $	0.32 \pm 0.11	$ & $	0.39	$ \\
\hline
\end{tabular}
\caption{Best fits to the Fundamental Plane, of the form $\log L_{\rm 1.4} = a_{\rm OIII} \log L_{\rm OIII} + a_{\rm mass} \log M + c$ where mass is either halo or black hole mass. Compact sources show less scatter than extended sources. There are too few extended HERGs (19) for meaningful fits; note that the extended HERGs also introduce large uncertainties into the overall HERG fits.}
\label{tab:fit_params}
\end{table}

\section{Discussion}
\label{sec:discussion}


\subsection{Dimensionless Fundamental Plane}
\label{sec:dimless_FP}

\citet{Daly16} presented a novel approach to non-dimensionalising the Fundamental Plane analysis, plotting the ratio $Q_{\rm jet} / L_{\rm bol}$ as a function of $L_{\rm bol} / L_{\rm Edd}$ for a sample of powerful classical double radio sources. Using a sample of high and low-excitation powerful radio galaxies, \citet{Daly16} argued that there appears to be a trend to lower jet production efficiencies (i.e. lower $Q_{\rm jet} / L_{\rm bol}$ values) with increasing accretion efficiency. \citet{XieYuan17} performed a similar analysis using radio and X-ray luminosities, and argued for a change in radio-loudness at the lowest accretion rates.

\begin{figure*}
	\includegraphics[width=0.68\columnwidth]{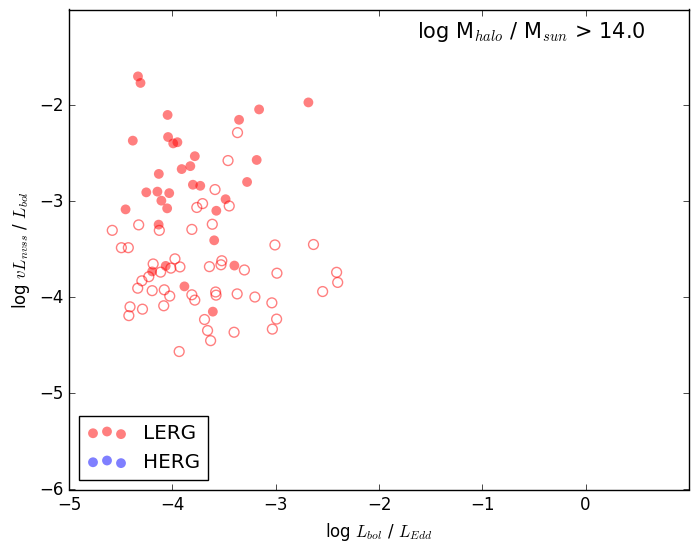}
	\includegraphics[width=0.68\columnwidth]{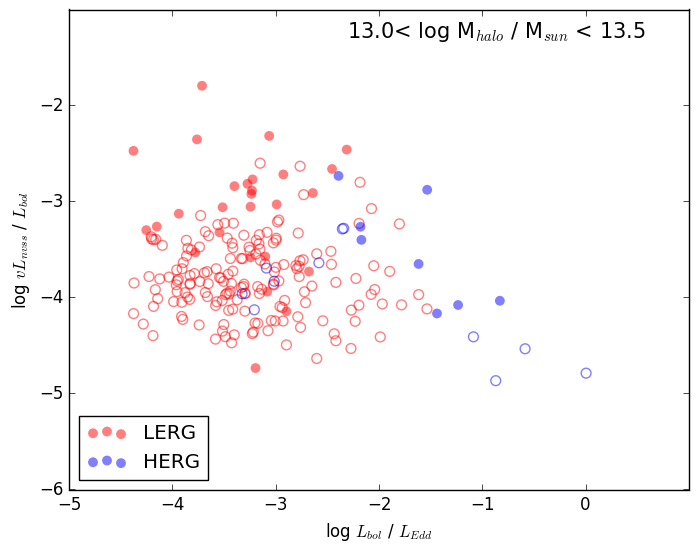}
	\includegraphics[width=0.68\columnwidth]{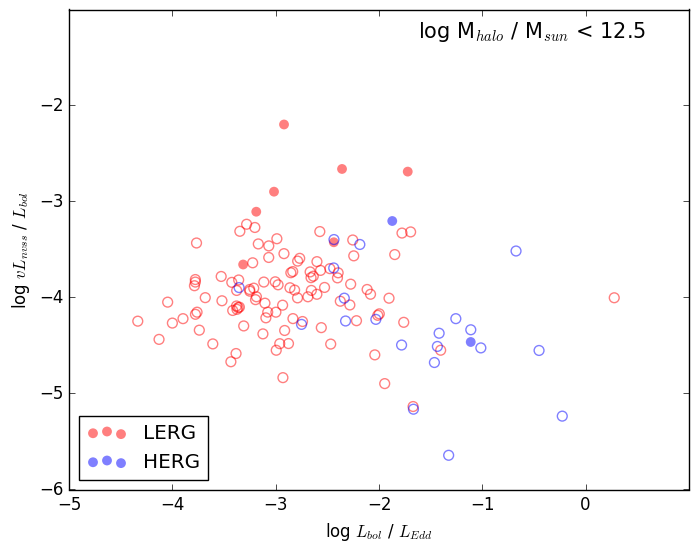}
	\caption{Relationship between radio loudness and dimensionless accretion rate, for different halo masses. Open circles again represent compact sources, closed circles are extended objects. Massive haloes host a higher fraction of both low accretion rate black holes, and extended radio AGN.}
	\label{fig:dimless_jetEfficiency}
\end{figure*}

In Figure~\ref{fig:dimless_jetEfficiency}, we plot the ratio of radio to bolometric luminosity as a function of dimensionless accretion rate\footnote{Using jet generation efficiency instead of radio luminosity, $\log \left( Q_{\rm jet} / L_{\rm bol} \right) = \frac{6}{5+p} \log L_{\rm radio} - \log L_{\rm bol} + {\rm const}$ (Equation~\ref{eqn:extended_radioBol_scaling}), yields similar results.}. We recover the result of \citet{Daly16} and \citet{XieYuan17} that lower accretion rate systems are more radio-loud. Our work also shows that environment is a hidden variable in these relations: massive haloes host a higher fraction of low accretion rate (in Eddington units) systems. Moreover, as Figures~\ref{fig:dimless_jetEfficiency} and \ref{fig:Lnvss_vs_Loiii_complete} show, massive haloes are also more likely to host extended, luminous radio sources.

\subsection{Implications of the environmental dependence of radio emission}
\label{sec:invisibleLobes}

The majority of sources in our sample appear compact, consistent with the earlier results of \citet{BestEA05,SAAR08,BaldiEA15}. Using the definition of Section~\ref{sec:results}, only 18 of 594 compact LERGs (our largest sub-sample) have significant excess of diffuse emission, $S_{\rm NVSS} / S_{\rm FIRST} > 5$. Relaxing the compactness criterion to $S_{\rm NVSS} / S_{\rm FIRST} > 1.38$ \citep{SinghEA15} increases the number of sources with extended emission to 170. This fraction is mass-dependent, with $24 \pm 3$\% of unresolved AGN in low-mass ($M_{\rm halo}<10^{13} M_\odot / h$) haloes having extended emission, compared with $33 \pm 4$\% in high-mass haloes ($M_{\rm halo}>3 \times 10^{13} M_\odot / h$). A possible interpretation of this compact population is that the compact jets are not young or frustrated, but instead have low-surface brightness radio lobes which are too faint for detection, similar to the subsample of compact, VLBI-detected AGN in poor environments studied by \citet{ShabalaEA17}. Extended low surface brightness radio emission is naturally expected in steep environments where the radio jets cannot be collimated to form classical Fanaroff-Riley extended structures. In this scenario, only emission on scales comparable to the galactic disk -- where the jet has a sufficient working surface -- will be seen. Free-free absorption by the multi-phase interstellar medium will produce characteristic GHz-peaked spectra on these kpc-scales, potentially explaining the overabundance of observed low-luminosity, compact, flat-spectrum sources \citep{WhittamEA17}.

\citet{YuanWang12} found a broad relation between core and lobe radio luminosities in both radio galaxy and quasar samples, with a (logarithmic) slope of less than unity; in other words, faint radio sources have a higher extended-to-core luminosity ratio. These low-luminosity jets tend to be LERGs in dense environments \citep{BestHeckman12,SabaterEA13}, and hence the shallow slope of the \citet{YuanWang12} relation is consistent with our results above.

The possibility of lobes contributing to detected radio emission has two potentially important implications. As discussed in Section~\ref{sec:theory}, compact and extended sources have different jet power -- radio luminosity scalings; misinterpreting lobe-dominated emission will therefore tend to overestimate jet kinetic powers, with implications for galaxy formation models \citep{RaoufEA17}. Moreover, the conversion from total to core luminosity \citep[e.g. using the relations of ][]{YuanWang12} must take environment into account. The second effect will be on using the Fundamental Plane relations to estimate black hole masses, as has been recently done for intermediate \citep{GultekinEA14,KoliopanosEA17} and ultra-massive \citep{HlavacekLarrondoEA12} black hole populations. Here, any lobe radio emission will lead to overestimates in black hole mass.
Observations sensitive to low-surface brightness emission, such as the MWA GLEAM survey \citep{HurleyWalkerEA17}, should soon quantify such selection effects and answer the question of whether the majority of radio AGN sources are genuinely compact, or their full extent is simply invisible to conventional interferometers at gigahertz frequencies. 


\section{Conclusions}
\label{sec:conclusions}

This short paper considers the environmental dependence on the optical Fundamental Plane of black hole activity. SDSS group catalogues were used to quantify environments around a sample of low-redshift ($z<0.2$) AGN. [O\,{\textsc{iii}}]  line luminosities are used the estimate the AGN bolometric luminosities, and VLA FIRST and NVSS surveys to quantify the 1.4 GHz radio continuum emission associated with the AGN on two spatial scales. The main result of this work is that mass of the AGN host halo is a hidden variable in the Fundamental Plane relations. Specifically:

\begin{itemize}
\item High and low-excitation radio galaxies occupy different mass haloes. These AGN follow similar jet power -- radio luminosity relations, but with different normalizations due to their different jet production efficiencies. When these populations are considered together, the preference of low-excitation radio AGN towards massive haloes naturally introduces a mass-dependent ``tilt'' in the AGN Fundamental Plane.

\item Even at fixed accretion mode, the fraction of compact and extended radio AGN depends on halo mass. Massive haloes are more likely to host AGN with substantial lobe, rather than jet, emission. The relationship between lobe radio luminosity and jet power is similar to that involving jet luminosity, but again with a different normalization. Lobe contribution to AGN radio luminosity also contributes to the mass-dependent tilt in the FP.

\item Scatter in the FP relations is reduced significantly when AGNs in different accretion modes, and with different jet/lobe contributions to radio emission, are treated separately. For compact low-excitation AGN, our largest sub-sample, scatter in the FP relations reduces by over 40 percent, and the halo mass dependence becomes statistically insignificant at the $2\sigma$ level. 
\end{itemize}

Similar results are found when black hole mass (estimated using stellar velocity dispersions) is used instead of halo mass. This is expected due to the well-known correlation between these two parameters, and suggests something rather important: our results above are consistent with a scenario in which the role of black hole mass in the observed optical Fundamental Plane of black hole activity is simply as a proxy for the real variable, namely the environment in which the AGN resides.



\section*{Acknowledgements}

I am grateful to Ivy Wong, Simon Ellingsen, Leith Godfrey and Martin Krause for many interesting conversations. I thank the anonymous referee for a constructive report which greatly improved the paper, particularly aspects related to stellar mass black holes. This work was partly funded by an Australian Research Council Early Career Fellowship (DE130101399).

\bibliographystyle{mnras}
\bibliography{FPletter_bibliography}


\bsp	
\label{lastpage}
\end{document}